\documentclass{elsart}
\usepackage{amsmath}

\newcommand{\tg}{\tilde{g}}
\newcommand{\tR}{\tilde{R}}
\newcommand{\Ham}{\mathcal{H}}
\newcommand{\G}{\mathcal{G}}
\providecommand{\e}{}
\renewcommand{\e}[1]{\mathrm{e}^{{#1}}}
\newcommand{\Hrest}{\Ham_{\text{rem}}} 
\newcommand{\Anom}{\mathcal{A}}
\newcommand{\Sren}{S_{\text{ren}}}
\newcommand{\hp}{\hat{\phi}}
\newcommand{\cp}{\check{\phi}}
\newcommand{\M}{\mathcal{M}}

\begin{document}

\begin{frontmatter}

\title{Holographic Renormalization and Ward Identities with the 
Hamilton--Jacobi Method\thanksref{prep}} 

\thanks[prep]{Preprint numbers: hep-th/0205061, QMUL--PH--02--11,
  DSF--11--2002} 

\author[QM]{Dario Martelli}
\ead{d.martelli@qmul.ac.uk} 
and
\author[NA]{Wolfgang M\"uck}
\ead{mueck@na.infn.it}

\address[QM]{Queen Mary College, University of London,
  Mile End Road, London E1 4NS, UK}

\address[NA]{Dipartimento di Scienze Fisiche, 
  Universit\`a di Napoli ``Federico II'',
  Via Cintia, 80126 Napoli, Italy}

\begin{abstract}
A systematic procedure for performing holographic renormalization,
which makes use of the Hamilton--Jacobi method, is proposed and
applied to a bulk theory of gravity interacting with a scalar
field and a $U(1)$ gauge field in the St\"{u}ckelberg formalism. 
We describe how the power
divergences are obtained as solutions of a set of ``descent
equations'' stemming from the radial Hamiltonian constraint of the theory. 
In addition, we isolate the logarithmic divergences, which are closely 
related to anomalies. The method allows to
determine also the exact one-point functions of the dual field theory.
Using the other Hamiltonian constraints of the bulk theory,
we derive the Ward identities for diffeomorphisms and gauge invariance.
In particular, we demonstrate the breaking of $U(1)_R$ current 
conservation, recovering the holographic chiral anomaly recently discussed in 
\texttt{hep-th/0112119} and \texttt{hep-th/0202056}.    
\end{abstract}


\end{frontmatter}


\section{Introduction and Summary}
\label{intro}
During the past four years, we have learned much from the AdS/CFT
correspondence \cite{Maldacena98,Gubser98-1,Witten98-1} about
conformal field theories (CFTs). The basic notion of the AdS/CFT
correspondence is that a gravitational theory (such as String Theory) 
living on $(d+1)$-dimensional
anti-de Sitter (AdS) space (the bulk space) is dual to a CFT living on its
conformal boundary. More generally, asymptotically AdS domain wall
solutions of $(d+1)$-dimensional gravity coupled to certain matter
fields are the duals of deformations of CFTs either by the addition of
relevant operators to the CFT Lagrangian or by the choice of a
non-conformal ground state involving vev's of certain operators. Using
a precise recipe one is able to obtain the correlation functions of
these (deformed) CFTs from the dynamics of the bulk theory. The most
important quantity is the on-shell action of the bulk theory, which,
after suitable regularization and renormalization, is identified with
the generating functional of the boundary field theory. 
For a recent exposition of the AdS/CFT correspondence including an
extensive list of references, we refer the reader to the lecture notes
\cite{dHoker02a}.

The occurrence
of divergences in the bulk on-shell action was noted already in the
earliest AdS/CFT calculations
\cite{Gubser98-1,Witten98-1,Mueck98-1,Freedman98-1}. The first
divergence to be explicitly removed from the on-shell action by
adding a counterterm was the boundary volume divergence in the context
of pure gravity on an AdS background \cite{Liu98-1}. Soon after, and
still for the case of pure gravity on an AdS background, the general
structure of the divergent terms and the relation of the logarithmic
divergence to the conformal anomaly of the boundary CFT was discussed
in \cite{Henningson98-2}. The addition of counterterms
also provided a way to give a meaning to the notion of energy in
asymptotically AdS spaces in terms of the renormalized Brown-York
stress energy tensor \cite{Brown93} without
the need of reference spaces
\cite{Balasubramanian99-2,Myers99-1,Emparan99-1}\footnote{For example,
compare the treatments of the AdS-Schwarzschild black hole using
pure AdS as reference space \cite{Witten98-2} and using counterterms
\cite{Cappiello01}.} 

A systematic development of holographic renormalization
for bulk gravity coupled to scalar fields was first given in
\cite{deHaro00a}. This method, which we shall refer to as the
\emph{standard method} of holographic renormalization, involves the
cancellation of all cut-off related divergences from the bulk on-shell
action by the addition of counterterms on a cut-off boundary hypersurface
and the subsequent removal of the cut-off. The counterterms removing
power divergences are fully covariant expressions of fields living at
the cut-off boundary, whereas logarithmic counterterms depend also explicitly
on the cut-off. This dependene breaks some bulk diffeomorphisms and
yields, \emph{e.g.}, the trace anomaly \cite{Henningson98-2}. 
Most recently, the standard method was summarized very clearly
by Bianchi, Freedman and Skenderis (BFS) \cite{Bianchi01b} (see also
the lecture notes by Skenderis \cite{Skenderis02a}) 
and applied to domain wall bulk geometries dual to deformed CFTs,
where also bulk gauge fields were considered. 
Holographic renormalization has been used in the
calculation of two-point functions in deformed CFTs \cite{Bianchi01}.

In a parallel development, de~Boer, Verlinde and Verlinde (dBVV)
\cite{deBoer00a} advocated the Hamilton-Jacobi (HJ) approach 
in order to separate terms in the bulk on-shell action, which can be
written as local functionals of the boundary data. The remaining,
presumably non-local, expression was identified with the generating
functional of a boundary field theory. This approach does not
correspond to the standard one, in particular, because dBVV's boundary
field theory lives on the cut-off boundary, and because the generating
functional still contains logarithmic divergences. Nevertheless, 
dBVV's method remains intriguing for its simplicity---it yields formally 
the correct gravitational anomalies and provides a remarkably simple
bulk description of the renormalization group flow in deformed CFTs.
A first attempt to use the Hamilton--Jacobi method for performing holographic 
renormalization appeared in \cite{Kalkkinen01b}, where 
a list of references to further studies of dBVV's method can be found. 

Despite its simplicity, the use of dBVV's method to perform
holographic renormalization does 
no appear to be very popular. The main drawbacks in its original
formulation seem to be the non-uniqueness of the solutions for the
local terms to be used as counterterms and the failure to obtain the
logarithmic counterterms. 
It is our intention in this paper to overcome these difficulties by
explaining exactly how the ambiguities can be removed from dBVV's
method and how logarithmic counterterms are obtained. The main steps
of our approach, which we shall call the \emph{HJ method} of
holographic renormalization, will be as follows. 
In the first step, we shall use dBVV's method to determine the
counterterms for the power divergences. Ambiguities in 
the descent equations are removed at the top level by comparison with
simple and known free field calculations. In the second step, we
continue to analyze the hamiltonian constraint with two
important results. First, we prove that all power 
divergences have been removed by the first step, and, second, we
obtain explicitly the logarithmic divergence and remove it. 
Hence, we prove that the HJ method qualifies as a consistent 
procedure for performing holographic renormalization. 
We point out that the difference between the 
standard method and the HJ method does not stem from using different
renormalization schemes, but regards only the procedure in which the
(exactly identical) counterterms are obtained. In both methods a
choice of renormalization scheme is reflected in the possibility of adding
finite counterterms. 

A very interesting aspect of the HJ formalism is the way 
holographic Ward identities emerge. In fact, local symmetries of the bulk
theory are encoded in hamiltonian constraints for the boundary data.
Exploiting the fact that the on-shell action is identified with 
the field theory generating functional, these can be naturally interpreted   
as Ward identities \cite{Corley00,Kalkkinen00}. The hamiltonian 
constraint associated with radial evolution can be seen as the Ward identity
for conformal symmetry, whose anomalous contributions are related 
to the logarithmic divergences of the on-shell action 
\cite{Henningson98-2}. Similarly, we will see that logarithmic terms
are also responsible for an anomalous term in the current Ward identity.

Let us give an outline of the rest of the paper and summarize our results.
In Sec.~\ref{holrenorm} we shall briefly review the standard method of 
holographic renormalization in order to familiarize the reader with
the issue and the complexity of the method, and to
allow a direct comparison with the approach that we will propose. 
We will then turn to the HJ method with the purpose of performing the
holographic renormalization. The method will be compared to the
standard method at appropriate points throughout the paper.    
For simplicity, in the first part of the paper
we shall confine our attention to a bulk gravity theory coupled to a
single scalar field. In Sec.~\ref{regren} we focus on
the procedure of regularization and renormalization of the bulk on-shell
action. This is carried out solving the hamiltonian constraint for
radial evolution in a recursive fashion. Following dBVV, we start with
a general ansatz for the leading local terms of the on-shell action,
which should be covariant  
and contain all power divergences. The constraint will then fix all of 
them, provided some little input is given for the lowest levels (Sec.\
\ref{powdiv}). As was
already discussed in \cite{deBoer00a,Kalkkinen01b}, there might be some
obstruction to the solution of the constraint, which is closely
related to the occurrence of a conformal anomaly.
We will make this observation more precise and relate the 
anomaly to the logarithmic divergences of the on-shell action, which we
can explicitly determine and remove. This is carried out in Sec.\
\ref{logardiv}, where we also prove that all divergences have been
removed.  

In Sec.~\ref{exact_1pt}, we shall discuss 
how to determine exact one-point functions in our method.  
By ``exact'' we mean one-point functions in the presence of sources turned on,
thus, in principle, encoding all higher point functions of the field theory.
This is one of the main results of BFS's approach, and any alternative
method should not fall short of it. Our results will also explicitly
demonstrate the scheme dependence of local terms in the exact
one-point functions. As a further application, we consider one-point
functions in bulk backgrounds, which are called holographic
renormalization group flows (Sec.~\ref{RG_1pt}). The difference
between operator and vev flows, which describe deformations of the
dual CFT by either operator insertion or a non-zero vev, respectively,
will become explicit. Moreover, the most natural choice of
counterterms will correspond to a supersymmetric renormalization
scheme.

In Sec.~\ref{vectors} we will include a $U(1)$
gauge field in our treatment, which should correspond to the
$R$-symmetry current of the dual field theory
\cite{anatomy,Bianchi01b}. A repetition of the arguments previously
explained will allow us to show how to solve for the  
on-shell action up to level four (with some simplifications)
and to obtain new contributions to the conformal anomaly. We will obtain 
some one-point functions for the new fields and, in particular, we will give
a full holographic derivation of the complete Ward identities, recovering 
also the chiral anomaly recently discussed in \cite{Bianchi01b},
\cite{KOW} and \cite{Bertolini:2002}.

\section{Holographic Renormalization -- Standard Approach}
\label{holrenorm}

To start, let us summarize the standard method of holographic renormalization 
as described by BFS \cite{Bianchi01b}. For later convenience, we shall
use a bulk metric of the form 
\begin{equation}
\label{intro:metric}
  ds^2 = dr^2 + g_{ij}(x,r) dx^i dx^j~,  
\end{equation}
where $i,j=1\ldots d$. The metric \eqref{intro:metric} is related to
the Fefferman-Graham metric for asymptotically AdS spaces
\cite{Fefferman85} by a simple change of variables. 
The asymptotic region is $r\to \infty$,
where the bulk approaches AdS spacetime with characteristic length $l$, 
\emph{i.e.},
\begin{equation}
\label{intro:AdS}
  g_{ij} (x,r) \to \e{2r/l} \hat{g}_{ij}(x)~.
\end{equation}
   
The standard method starts by solving the equations of motion in the
asymptotically AdS region, which yields the bulk fields in the form of
asymptotic series like\footnote{Although this form is not generic, it
  suffices to describe the standard method. The expansion
  \eqref{intro:asympt} will in general contain not only powers of
  $\e{r/l}$ and the logarithmic terms described later, because bulk
  interactions and generic values of the conformal dimensions will
  generate non-integer exponents. The literature on the standard
  method \cite{deHaro00a,Bianchi01b} starts with expansions of the
  form \eqref{intro:asympt}, because it covers most interesting cases,
  where conformal dimensions are integer or half-integer. However,
  there is no obstruction in principle to apply the method to other
  cases \cite{Skenderis_priv}.}
\begin{equation}
\label{intro:asympt}
\begin{split}
  F(x, r) &= \e{\lambda_1 r/l} \left[ f_0(x) + \e{r/l} f_1(x) 
  + \e{2r/l} f_2(x) +\cdots
  \right] \\
  &\quad + \e{\lambda_2 r/l} \left[ \tilde{f}_0(x) + 
  \e{r/l} \tilde{f}_1(x) + \e{2r/l} \tilde{f}_2(x) +\cdots \right]~,
\end{split}
\end{equation} 
consisting of the two independent asymptotic series solutions of the
equations of motion. These series expansions are similar to the
Fefferman--Graham expansion of the metric for asymptotically AdS spaces
\cite{Fefferman85} in pure gravity. 

The assumption that the bulk space be asymptotically AdS implies that
the fields are asymptotically free, and thus the coefficients
$\lambda_1$ and $\lambda_2$ are related to the asymptotic mass of the
field $F$, whereas $f_0$ and $\tilde{f_0}$ are independent integration
constants. Without loss of generality, we shall assume that
$\lambda_1>\lambda_2$, so that the first line in eqn.\
\eqref{intro:asympt} represents the leading series solution. Then,
$f_0$ is called the \emph{source} of $F$.  
The coefficients $f_1$, $f_2$, etc., and similarly, $\tilde{f}_1$,
$\tilde{f}_2$, etc., are obtained by recursive
analysis of the bulk field equations in the asymptotic region and 
depend locally on the $f_0$ and $\tilde{f}_0$ of all fields,
respectively. In the dual boundary field theory, the source $f_0$ is
the generating current of the operator that is dual to the bulk field
$F$, while $\tilde{f}_0$ is related to the vev of this operator (see
Sec.\ \eqref{exact_1pt}). 

For example, for the metric, $g_{ij}$, the coefficients are
$\lambda_1=2$ and $\lambda_2=2-d$. In the
case of a scalar field, $\phi$, of bulk mass $m$, they are 
$\lambda_1=-(d/2-\lambda)$ and $\lambda_2=-(d/2+\lambda)$, where $\lambda
= \sqrt{d^2/4+m^2l^2}$. The bulk field $\phi$ is the dual of a scalar
operator with conformal dimension $\Delta =d/2+\lambda$. 
We shall consider only matter fields dual to relevant operators, for which
$\lambda_1<0$, so that they tend to zero in the asymptotic
region.\footnote{One could also include \emph{marginal} operators,
  which have $\lambda_1=0$ \cite{Skenderis02a}.} 
For even values of $d$, there are logarithmic terms in the leading
series for the metric (proportional not
to an exponential of $r$, but to $r$), starting with $r\times \exp[(2-d)r/l]$. 
Similarly, logarithmic terms occur in the leading series of
scalar fields for integer $\lambda$. 

Having obtained the asymptotic form of the bulk fields, one proceeds
to calculate the regulated on-shell action, which is the on-shell
action for a bulk spacetime with a cut-off boundary at $r=\rho$,
\begin{equation}
\label{intro:Sreg}
  S_{\text{reg}} [f_0, \rho] = \int d^d x \sqrt{g_0}
  \left[\e{\nu_0\rho/l} a_0(f_0) + \e{\nu_1 \rho/l} a_1(f_0)
   +\cdots + \frac{\rho}l
  \bar{a}(f_0) + O(1) \right]~, 
\end{equation} 
where the $\nu_k$ are positive numbers. There are a finite number of
divergent terms in the limit $\rho\to \infty$, which comprise the
power divergences with coefficients $a_{k}$ and a
logarithmic divergence with a coefficient $\bar{a}$ that is related
to the appearance of logarithmic terms in the expansion
\eqref{intro:asympt}. The coefficients $a_{k}$ and $\bar{a}$ are 
local functions of the coefficients $f_0$ of the leading series of the bulk
fields and of their derivatives. 

The method of holographic renormalization proceeds now as
follows. First, the series \eqref{intro:asympt} are inverted for all
fields and solved for the sources $f_0(x)$ as functionals of the
$F(x,r)$. Second, the $f_0$ thus
obtained are substituted into eqn.~\eqref{intro:Sreg}, which yields
the divergent terms of the on-shell action in terms of the fields
$F(x,\rho)$ living at the cut-off boundary. Finally, these terms are
subtracted from $S_{\text{reg}}$ (using counterterms), and in the 
remaining expression the limit $\rho\to\infty$ is 
taken. The finite result (expressed again as a functional of the
sources $f_0$) is the renormalized on-shell action to be identified
with the generating functional of the boundary field theory. 

BFS's method is clear and rigorous. The four steps of the 
procedure---the asymptotic expansion of the fields, the recursive
determination of the coefficients, the inversion of the asymptotic
series and the 
substitution of the inverted series into the on-shell action---form an
algorithm that uniquely yields all counterterms needed to cancel the  
$\rho\to \infty$ divergences of the bulk on-shell action. 
Further finite counterterms could be added and correspond to a certain
choice of renormalization scheme. One should notice the
importance of using covariant counterterms, hence the need to invert
the asymptotic series solutions of the fields. Covariance of the
counterterms implies the validity of the Ward identity of
diffeomorphisms of the boundary field theory. In fact, the
counterterms, when expressed in terms of $f_0$ and $\tilde{f}_0$,
contain finite terms that affect the renormalized on-shell action and
do not correspond to a choice of renormalization scheme, \emph{i.e.},
they contribute to the physical results. 

It is an interesting question whether there exists a method, which
determines the counterterms directly at the cut-off boundary without
performing the asymptotic expansion and its inversion. Such a method is
provided by the Hamilton-Jacobi approach, which we shall turn to in the
next section.

\section{Holographic Renormalization -- Hamilton-Jacobi Approach}
\label{regren}
As mentioned in the previous sections, we propose to use the
HJ method advocated for the AdS/CFT correspondence 
by dBVV in order to determine the counterterms. In this section, we
shall describe in detail how the counterterms are obtained, starting 
with the power divergences in Sec.~\ref{powdiv} and discussing the
logarithmic divergences in Sec.~\ref{logardiv}. The relation
between logarithmic divergences and conformal anomalies will become
explicit, a fact that has been used by Henningson and Skenderis
\cite{Henningson98-2} in the derivation of the conformal anomaly.

In order to keep the presentation simple and to concentrate on the
main steps of the method, we consider in this section 
the action of gravity coupled to a single scalar
field\footnote{Details of our notation are given in the 
appendix. For the more general case of several scalars, see
\cite{Kalkkinen01b}.}
\begin{equation}
\label{action}
  S = \int d^{d+1}x \sqrt{\tg} \left[ -\frac14\tR + \frac12
  \tg^{\mu\nu} \partial_\mu \phi \partial_\nu \phi +V(\phi)  \right] 
  + \frac12 \int d^d x \sqrt{g}\, H~.
\end{equation}
The necessary formulae for the HJ method are summarized in the
appendix, and the vector sector should be omitted 
here. We shall include the gauge fields in Sec.~\ref{vectors}.

Before plunging into the details, let us briefly compare the
HJ approach with the standard method of holographic
renormalization. In the HJ approach, the analysis is
carried out directly at the cut-off boundary, so that an asymptotic
expansion and its inversion are not necessary, except for the leading
order term determined by the behaviour of the free fields.  
The recursive determination of the series coefficients from
the field equations is replaced by the recursive solution of a set of
descent equations. In order to determine where to stop (if the system
does not break down by itself), a simple power counting using the
leading behaviour of the fields is sufficient. 
The descent equations are derived from one equation
only, which is the hamiltonian constraint, using the independence of
the boundary conditions in order to split the constraint into
functionally independent terms. One should expect
ambiguities, because this equation does not contain all information
about the field dynamics. In 
fact, the descent equations do not yield a unique solution for the
counterterms, but, as we shall see, the ambiguities appear right at the top
of the descent equations and can be easily removed with little input
from the (known) results for free fields. 
The subsequent analysis of the descent equations
yields a unique answer for the divergent counterterms. As a very nice
feature, the other constraint equations translate into Ward identities of
the dual field theory.

\subsection{Power Divergences}
\label{powdiv}
In this section, we shall briefly outline how the power divergences
of the on-shell action are obtained using the method of dBVV. 
In order to separate the power divergences, the on-shell action is
split up as
\begin{equation}
\label{split_S}
  S = S_{[0]} + S_{[2]} + \cdots + S_{[2n]} + \Gamma~,
\end{equation}
where we have denoted power divergent terms by $S_{[2k]}$, $k=0\ldots n$. 
According to eqn.~\eqref{pi_S}, the momentum $\pi$ naturally splits
into 
\begin{equation}
\label{split_pi}
  \pi = \pi_{[0]} + \pi_{[2]} + \cdots + \pi_{[2n]} + \pi_\Gamma~,
\end{equation}
and similarly for $q^i_j$. 
We shall make the following premises regarding the counterterms
$S_{[2k]}$. 
\begin{enumerate}
\item The counterterms are covariant (and gauge invariant, if vector
  fields are involved) local expressions in terms of the metric
  $g_{ij}$ and the field $\phi$ at the cut-off boundary as well as
  their derivatives.  
\item The term $S_{[2k]}$ contains exactly $k$ inverse metrics. (This
  is an arbitrary assumption, but turns out to be very useful
  for the bookkeeping.) 
\item The counterterms should completely contain the power divergences of $S$.
\item The counterterms must be universal, \emph{i.e.}, they must
  contain the power divergences of $S$ for \emph{any} asymptotically AdS
  solution of the bulk equations of motion. 
\end{enumerate}

The premises 1 and 2 imply that we can write 
\begin{align}
\label{S0}
  S_{[0]} &= \int d^dx \sqrt{g}\, U(\phi)~,\\
\label{S2}
  S_{[2]} &= \int d^dx \sqrt{g}\, \left[ \frac12 M(\phi) g^{ij} \partial_i
  \phi \partial_j \phi + \Phi(\phi) R \right]~, \text{ etc.}
\end{align}
The number $n$ of power divergent terms is dictated by premise 3:
By power counting using the leading asymptotic behaviour
of the fields for $r\to \infty$ [cf.~eqn.~\eqref{intro:asympt}], one
can determine which covariant boundary integrals are generically
divergent. All of these have to be included. 
For example, an $S_{[2k]}$ without derivatives of matter fields
behaves generically like $\exp[(d-2k)\rho/l]$ from the metric contributions. 
Thus, we must include all of these with $k<d/2$.
In eqn.~\eqref{split_S}, $\Gamma$ may contain a
logarithmic divergence, but 
shall be regarded as finite for the power counting. Finite local terms
can be arbitrarily shifted between $\Gamma$ and the counterterms. This
reflects the usual ambiguity of choosing a renormalization scheme.

The main point of dBVV's method is to analyze the Hamiltonian
constraint [eqn.~\eqref{con_H}] by splitting it into a derivative 
expansion,\footnote{Such kinds of expansions were proposed previously in
  the literature as a method for solving the Hamilton--Jacobi
  equations of General Relativity in the presence of some matter
  fields \cite{Parry:1994mw,Darian:1998mp}.}
\begin{equation}
\label{split_H}
\Ham = \Ham_{[0]} + \Ham_{[2]} +\cdots +\Ham_{[4n]} +\Ham_{\Gamma} =
0~,
\end{equation}
where $\Ham_{[2k]}$ denotes those terms in $\Ham$ that stem only from the
counterterms and contain a total of $k$ inverse metrics. Notice that
eqn.\ \eqref{split_H} only re-writes the hamiltonian $\Ham$ in a form,
which makes the contributions from the various counterterms explicit.
By construction, a certain number of $\Ham_{[2k]}$ are asymptotically
stronger than $\Ham_{\Gamma}$, and these can be found by simple power
counting. One should now try to solve the constraint
separately for these (stronger) $\Ham_{[2k]}$, starting with
$\Ham_{[0]}$, which leads to a system of descent equations. 
This approach is justified by the universality premise.

Let us demonstrate the procedure by considering 
the level 0 and level 2 terms of the Hamiltonian. For more details,
see \cite{deBoer00a,Kalkkinen01b}. The level 0 descent equation
is given by 
\begin{equation}
\label{H0}
  \Ham_{[0]} = -\frac{d}{d-1} U^2 +\frac12 \left(U'\right)^2 - V = 0~.
\end{equation}
This equation is regarded as a functional equation for $U(\phi)$, where
$V(\phi)$ is given. Solving eqn.~\eqref{H0} is impossible in closed form for a
generic potential $V(\phi)$  (see \cite{Campos00,Martelli02a} for a
numerical analysis).   
However, we are interested in removing the $\rho\to \infty$
divergences, and thus an analysis in the asymptotic region is
sufficient. In fact, because the leading behaviour of a relevant scalar
is $\phi(x,r) \sim \phi_0(x)\e{\lambda_1 r}$ with $\lambda_1<0$
[cf.\ eqn.\ \eqref{intro:asympt}], for
\emph{any finite, but otherwise arbitrary} value of the source $\phi_0$,
we can choose the cut-off $\rho$ sufficiently large such as to bring
the Dirichlet value $\phi(x,\rho)$ within an arbitrarily 
small neighbourhood of the fixed point, $\phi=0$. The same is true for any
number of relevant matter fields. Hence, it is sufficient to solve
$\Ham_{[0]}=0$ close the fixed point. For this purpose, we expand it into a
Taylor series.\footnote{Note that this is not equivalent to a large
  $\rho$ expansion of the descent equations, because it does not take
  into account the different large $\rho$ behaviours of different
  scalars \cite{Skenderis_priv}.}
Thus, we write $V(\phi)$ as
\begin{equation}
\label{V}
  V = -\frac{d(d-1)}{4l^2} +\frac12 m^2 \phi^2 +\frac1{3!} v_{3}
  \phi^3 + \frac1{4!}v_4 \phi^4 + O(\phi^5)~,
\end{equation}
where the constant part represents the negative cosmological constant,
and $m$ is the bulk mass parameter for $\phi$. Of course, this assumes
that the potential $V$ can be expanded in a Taylor series, which we
shall assume to be true. 
Similarly, we expand $U$ as 
\begin{equation}
\label{U}
  U = u_0 + u_1 \phi + \frac12 u_2 \phi^2 + \frac1{3!} u_{3}
  \phi^3  + \frac1{4!}u_4 \phi^4 + O(\phi^5)~.
\end{equation}
Now, we obtain from eqn.~\eqref{H0} a set of coupled equations, 
from which the coefficients $u_k$ should be recursively determined. 
It turns out that there is no unique solution, but input 
from the (known) free field behaviour will fix the ambiguities.
First, we see that the action \eqref{action} does not contain
terms, which are linear in $\phi$. Thus, there cannot be divergent
terms linear in $\phi$, and we shall set 
\begin{equation}
\label{U1}
  u_1=0~.
\end{equation}
This solves the term of order $\phi$ of eqn.~\eqref{H0}. 
Then, we find from the constant term
\begin{equation}
\label{U0_2}
  (u_0)^2 = \frac{(d-1)^2}{4l^2}~. 
\end{equation}
Here, we must choose a sign, and in order to choose the correct one,
let us compare with a simple known case. Liu and
Tseytlin \cite{Liu98-1} determined this leading counterterm as
\begin{equation}
\label{U0}
  u_0 = -\frac{d-1}{2l}~,
\end{equation}
which cancels the volume divergence of the on-shell action for pure gravity
\cite{Liu98-1}. This is easy to verify for a bulk AdS spacetime with a flat
cut-off boundary. Next, the quadratic term of eqn.~\eqref{H0} yields 
\begin{equation}
\label{U2_2}
  (u_2)^2 +\frac{d}l u_2 -m^2 =0~, 
\end{equation}
where eqns.~\eqref{U1} and \eqref{U0} have been used.
Again, the solution is not unique, and we must resort to the free
scalar field in AdS background, whose on-shell action is 
\begin{equation}
\label{S_scalar}
  S = \frac12 \int d^dx \sqrt{g}\, \phi \partial_r \phi~,
\end{equation}
and whose leading behaviour (obtained from the free equation of
motion) is  
\begin{equation}
\label{scalar_asym}
  \phi(x,r) \sim \e{(-d/2+\lambda)r/l} \hat{\phi}(x)
\end{equation}
with $\lambda = \sqrt{d^2/4+m^2l^2}$. Hence, the leading behaviour of
the on-shell action \eqref{S_scalar} is 
\begin{equation}
\label{S_scalar_asym}
  S \sim \frac1{2l} \int d^dx \sqrt{g}\, \phi^2 \left(-\frac{d}2+\lambda
  \right)~.
\end{equation}
This tells us that the correct choice amongst the solutions of eqn.\
\eqref{U2_2} is 
\begin{equation}
\label{U2}
  u_2 = \frac1l \left(-\frac{d}2 +\lambda \right)~.
\end{equation}
Continuing with the cubic term of eqn.~\eqref{H0}, we obtain 
\begin{equation}
\label{v3}
  v_3 = \frac{u_3}l \left( 3\lambda -\frac{d}2 \right)~,
\end{equation}
so that, for $\lambda \neq d/6$, we uniquely find
\begin{equation}
\label{U3}
  u_3 = \frac{2l}{6\lambda -d} v_3~.
\end{equation}
In contrast, for $\lambda=d/6$, the coefficient $u_3$ remains
undetermined, and we have a remaining term, 
\begin{equation}
\label{Hrest_0}
  \Hrest = - \frac16 v_3 \phi^3 +O(\phi^4)~.
\end{equation}
We shall postpone the discussion of $\Hrest$ until Sec.\
\ref{logardiv}. The fact that $u_3$ remains undetermined is a sign of the
renormalization scheme dependence. In fact,
\begin{equation}
\label{scheme_0}
  \int d^dx \sqrt{g}\, \phi^3 \sim \e{dr/l}
  \left[\e{(-d/2+\lambda)r/l}\right]^3 = 1 \quad \text{for
  $\lambda=d/6$}~,
\end{equation}
so that the undetermined counterterm is finite. More generally, a
breakdown of the level 0 descent equation will occur, if 
\begin{equation}
\label{break0}
  \lambda = \frac{(k-2)d}{2k}
\end{equation}
for some integer $k>2$ \cite{Martelli02a}.
In this case, the coefficient $u_k$ remains
undetermined, and the boundary integral of $\phi^k$ is finite in the
$\rho\to\infty$ limit. A relation similar to eqn.\
\eqref{break0} can be found also at the higher levels. 

For later use, let us also consider the $\phi^4$
term of the constraint \eqref{H0}, which becomes
\begin{equation}
\label{v4}
  \frac{u_4}{24l}(4\lambda-d) -\frac{v_4}{24}-\frac{d}{4(d-1)l^2}
  \left(\frac{d}2-\lambda \right)^2  +\frac{l^2
  v_3^2}{2(6\lambda-d)^2} = 0
\end{equation}
after the coefficients $u_0$, $u_1$, $u_2$ and $u_3$ have been
inserted. Clearly, if $\lambda \neq d/4$, eqn.~\eqref{v4} determines
$u_4$ uniquely, but, if $\lambda=d/4$, the descent equation breaks and
there is a remainder
\begin{equation}
\label{Hrest_4}
  \Hrest = \left( -\frac{v_4}{24} -\frac{d^3}{64(d-1)l^2}
  +\frac{2l^2v_3^2}{d^2} \right) \phi^4 +O(\phi^5)~.
\end{equation}
It turns out that the relevant contribution to $\Hrest$ might be zero in 
particular cases. An example is the potential responsible for the GPPZ flow
\cite{Girardello99a}, where $d=4$ and $\lambda=d/4=1$. The breakdown
occurs for $k=4$, but we have $v_4=-8/l^2$ and $v_3=0$, so that there is no
$\phi^4$ contribution to $\Hrest$. 

It is useful to pause for a moment and reflect on some details of what
we have done so far. 
The idea for finding the power divergent terms is to write down
the most general set of covariant local counterterms up to the necessary level
and solve for them by asking that they satisfy the hamiltonian constraint for
arbitrary and independent values of boundary conditions (Dirichlet
values at the cut-off boundary) of the fields. One might object that
this assumes that all power divergences can be cancelled by covariant
local counterterms, whereas, in the standard method, this is shown by
explicit construction. At this stage of the procedure, we do not have
any proof that the procedure applied so far yields all power
divergences, but this will be proven in Sec.~\ref{logardiv}.
  
The origin of the ambiguities we have met becomes clear when
considering the hamilton equations of motion, \emph{e.g.}, 
eqns.\ \eqref{momenta_q} and \eqref{momenta_pi}. The choice of sign of
$u_0$ reflects the fact that one can choose both, $r\to\infty$ and
$r\to-\infty$, as the asymptotic region, and the choice in eqn.\ \eqref{U0}
corresponds to $r\to\infty$. Similarly, the two possible solutions of
eqn.\ \eqref{U2_2} describe, via eqn.\ \eqref{momenta_pi}, the two
independent asymptotic behaviours of the scalar field. The counterterm
must describe the leading one, because it is asymptotically strongest,
and this is ensured by the choice in eqn.\ \eqref{U2}. It is now
also clear that further ambiguities can appear neither for higher
powers of $\phi$ nor at the higher levels. In fact, the hamilton
equations of motion provide a one-to-one correspondence between the
first few terms of the leading asymptotic series and the
counterterms. Hence, since the leading asymptotic series solution is
unique for any given sources, there is exactly one set of universal
covariant counterterms.

Let us now turn to the level 2 equation, which reads
\begin{align}
\label{H2}
\Ham_{[2]} &=\left(-\frac{d-2}{d-1} UM + 4U \Phi''-
 \frac12 U' M' - \frac12 \right)\, \nabla_i\phi\nabla^i\phi\nonumber\\
   &+ \Big( 4U \Phi' - U' M \Big)\,\nabla_i\nabla^i\phi\nonumber\\
   &+ \left(-2 \frac{d-2}{d-1} \Phi U +U' \Phi' + \frac14 \right)\,R  = 0~.
\end{align}
By the universality premise, functionally independent terms must
vanish separately, and this yields the following three descent equations,
\begin{align}
\label{level2a}
  -\frac{d-2}{d-1} UM + 4U \Phi''- \frac12 U' M' &= \frac12~,\\
\label{level2b}
  4U \Phi' - U' M &=0~,\\
\label{level2c}
  2 \frac{d-2}{d-1} \Phi U -U' \Phi' &= \frac14~. 
\end{align}
These have been analyzed in \cite{Kalkkinen01b}, and the solutions are   
\begin{equation}
\label{Phi}
  \Phi = \Phi_0 + \Phi_1 \phi +\frac12 \Phi_2 \phi^2 +O(\phi^3)~,
\end{equation}
with 
\begin{align}
\label{Phi0}
  \Phi_0 &= -\frac{l}{4(d-2)} \quad \text{for $d\neq 2$}~,\\
\label{Phi1}
  \Phi_1 &=0~,\\
\label{Phi2}
  \Phi_2 &= -\frac{(d-2)(d-2\lambda)}{2(d-1)(\lambda-1)} \Phi_0 \quad \text{for
  $\lambda\neq 1$}~,
\end{align}
and 
\begin{equation}
\label{M}
  M = \frac{l}{2(\lambda-1)} +O(\phi) \quad \text{for $\lambda \neq 1$}~.
\end{equation}
For $d=2$ and $\lambda=1$ the coefficients $\Phi_0$, $\Phi_2$ and
$M_0$ remain undetermined, respectively, leaving a remainder
\begin{equation}
\label{Hrest2}
  \Hrest = \begin{cases}
   -\frac12 \nabla^i \phi \nabla_i \phi - \frac{d-2}{8(d-1)}
     \phi^2 R +O(\phi^3, R^2) & (\lambda=1)~,\\
   \frac14 R +O(\phi,R^2) & (d=2)~. 
  \end{cases}
\end{equation}

Similarly, an analysis of the level 4 descent
equations has been carried out in \cite{Kalkkinen01b}. A breakdown at
level 4 occurs, if $d=4$ or $\lambda=2$. For $d=4$, the unresolved
remainder is 
\begin{equation}
\label{Hrest4}
  \Hrest = -\frac{l^2}{16} \left( \frac13 R^2 -R_{ij} R^{ij} \right)
   +O(R^3)~.
\end{equation}

The method can be extended recursively to any desired level, of course
with a rapidly increasing amount of complexity. The
highest level necessary to cancel the power divergent terms in $S$ is
found by power counting. 
As discussed above, the method yields all covariant boundary terms in the
on-shell action, which are divergent by power counting. These are
interpreted as counterterms to be added, although it has not yet been
shown that all power divergences of $S$ are obtained. It also gives
information on possible scheme dependent finite terms, and one may
choose to add also irrelevant counterterms. We shall turn to the
analysis of $\Gamma$ in the next section, where it is shown that it is
either finite or logarithmically divergent, which is indistinguishable
from finite behaviour in the power counting.

\subsection{Logarithmic Divergences and Anomalies}
\label{logardiv}
We have seen in Sec.~\ref{powdiv} that a zero coefficient in
front of one of the unknowns of the descent equations 
leads to a breakdown of the recursion and generically leaves behind
an unresolved remainder, $\Hrest$. We shall discuss the meaning of
$\Hrest$ now. In particular, we will show that it leads to a logarithmic
divergence of $\Gamma$ and to a conformal anomaly in the boundary
field theory. If we find that $\Hrest=0$ despite the breakdown, as 
it happens at level $0$ in the case of the GPPZ
flow potential, then there is no contribution to the anomaly.  

The argument continues with the analysis of the Hamiltonian
constraint, which, by solving the descent equations up to the
breakdown, has been reduced to  
\begin{equation}
\label{anom_eq}
  \Hrest + \Ham_\Gamma = 0~.
\end{equation}
Moreover, we can express $\Ham_\Gamma$ to leading order as
\begin{align}
\notag
  \Ham_\Gamma &= 8 q^{ij}_\Gamma q_{[0] ij} -\frac8{d-1} q_\Gamma
  q_{[0]} + \pi_\Gamma \pi_{[0]} +\cdots \\
\notag &= -\frac{4U}{d-1} q_\Gamma + U' \pi_\Gamma + \cdots \\
\label{Hgamma}
  &= \frac1l \left[ \frac{2g_{ij}}{\sqrt{g}} \frac{\delta
  \Gamma}{\delta g_{ij}} + \left(-\frac{d}2 +\lambda \right) \phi
  \frac1{\sqrt{g}} \frac{\delta \Gamma}{\delta \phi} \right] +\cdots~.
\end{align}
Here, we have used the results for $U$ from the previous section, 
and the ellipses stand for terms that are irrelevant in the
$\rho\to\infty$ limit. Defining (remember $\Delta =d/2+\lambda$)
\begin{equation}
\label{Adef}
  \Anom = \frac{2g_{ij}}{\sqrt{g}} \frac{\delta
  \Gamma}{\delta g_{ij}} + \left(\Delta-d \right) \phi
   \frac1{\sqrt{g}} \frac{\delta \Gamma}{\delta \phi}~,
\end{equation}
we find from eqns.~\eqref{anom_eq} and \eqref{Hgamma} 
\begin{equation}
\label{A}
  \Anom = - l \Hrest+\cdots~.
\end{equation}
The quantity $\Anom$ is what dBVV call the conformal anomaly of the
boundary field theory. This identification would follow from eqn.\
\eqref{Adef}, if $\Gamma$ were the generating functional of the
boundary field theory. Formally, $\Anom$ has the form of an anomaly, as can
be seen in the particular cases.
However, this interpretation is not
correct. First, $\Gamma$ contains logarithmic divergences and is
therefore not the generating functional of the boundary field
theory. Second, $\Anom$ scales like $\e{-d\rho/l}$ for $\rho\to\infty$,
\emph{i.e.}, it has no finite limit, and, third, the boundary data
$g_{ij}$ and $\phi$ are not the (renormalized)
sources of the boundary field theory. 
We shall derive below that the true anomaly is
obtained from $\Anom$ by a simple rescaling and taking the
$\rho\to\infty$ limit. However, it is important to notice that the boundary
integral of $\Anom$ is finite for $\rho\to\infty$ and has the form of
an integrated anomaly, \emph{i.e.}, it is conformally invariant.   

But first, we shall evaluate the logarithmic divergence of $\Gamma$. 
The argument
uses the diffeomorphism invariance of $\Gamma$ and is similar to the
arguments used in the derivation of the conformal anomaly by
Henningson and Skenderis \cite{Henningson98-2}. 
$\Gamma$ is a functional of the boundary data
$g_{ij}$ and $\phi$ and depends explicitly on the value of the
cut-off, $\rho$. We shall write it as $\Gamma[g_{ij}(x,\rho),
\phi(x,\rho);\rho]$. By construction, $\Gamma$ is invariant
under bulk diffeomorphisms,
\begin{equation}
\label{Gamma_inv}
  \Gamma[g_{ij}(x,\rho),\phi(x,\rho);\rho] = 
  \Gamma[g'_{ij}(x',\rho'),\phi'(x',\rho');\rho']~.
\end{equation}
In particular, under the following change of variables,
\begin{equation}
\label{var_change}
  x' = x~,\qquad r' = r (1+\sigma)~,
\end{equation}
the variation of $\Gamma$ becomes to first order 
\begin{multline}
\label{Gamma_inv2}
  \Gamma[g_{ij}(x,\rho'),\phi(x,\rho');\rho'] - 
  \Gamma[g_{ij}(x,\rho),\phi(x,\rho);\rho] \\ 
  = \rho \sigma \int d^dx
  \left[ \partial_\rho g_{ij} \frac{\delta \Gamma}{\delta g_{ij}} + 
  \partial_\rho \phi \frac{\delta\Gamma}{\delta\phi} \right]
  +\mathcal{O}(\sigma^2)~.
\end{multline}
Hence, after expressing $\partial_\rho g_{ij}$ and $\partial_\rho\phi$ to
leading order, we find
\begin{equation} 
\label{dGamma}
  \partial_\rho \Gamma = \frac1l \int d^d x \sqrt{g}\, \Anom +
  \cdots~,
\end{equation}
where the boundary values $g_{ij}$ and $\phi$ are kept fixed when
differentiating with respect to $\rho$ on the left hand side. 
Since the integral on the right hand side of eqn.~\eqref{dGamma} is
finite for $\rho\to\infty$, this equation implies
\begin{equation} 
\label{logdiv}
  \Gamma = \frac{\rho}l \int d^d x \sqrt{g}\, \Anom + \text{finite terms}~,
\end{equation}
which is the logarithmic divergence of $\Gamma$. Eqn.\
\eqref{logdiv} proves that $\Gamma$ does not contain power divergent
terms. In other words, we have found \emph{all} power divergent terms
of the on-shell action by the analysis in Sec.\ \ref{powdiv}. 

It is interesting to
ask why a breakdown of the descent equations is related to a logarithmic
divergence. In the standard method, one solves the equations of motion
in form of an asymptotic series. It is well known that, for certain
powers of the leading behaviour, the series form of the solution
will ``break down'', and one requires logarithms in order to continue
the recursion. In the HJ method, the descent equations determine the momenta
as a series of terms that locally depend on the fields, which, by
virtue of the hamilton equations 
of motion, corresponds to the leading asymptotic series solution. 
Hence, the descent equations break exactly at the point, where
the asymptotic series in the standard method would require a
logarithm, and this translates into a logarithmically divergent term
in the on-shell action.

Having found the logarithmic divergence, we define the
renormalized on-shell action as 
\begin{equation}
\label{S_ren}
  \Sren = \lim_{\rho\to\infty} \left( \Gamma + \rho \int d^d x
  \sqrt{g}\,\Hrest \right)~.
\end{equation}

Since $\Anom$ has the form of a conformal anomaly, but vanishes for
$\rho\to\infty$, it is natural to define a finite, rescaled quantity
(cf.~\cite{Kalkkinen01b}),
\begin{equation}
\label{Anom_def}
  \hat{\Anom} = \lim_{\rho\to\infty} \left( \e{d\rho/l} \Anom \right)~.
\end{equation}
$\hat{\Anom}$ is formally identical to $\Anom$ with $g_{ij}$ and $\phi$
replaced by $\hat{g}_{ij}$ and $\hat{\phi}$ (the coefficients of the
leading terms in the asymptotic expansions), respectively. 
Then, the expression 
\begin{equation}
\label{Anom_int}
  \int d^dx \sqrt{\hat{g}}\, \hat{\Anom} = \lim_{\rho\to\infty} 
  \int d^dx \sqrt{g}\, \Anom~
\end{equation}
is conformally invariant, and we confirm that $\hat{\Anom}$ is the
conformal anomaly of the boundary field theory,\footnote{We take
  $-\hp$ as the source coupling to the operator $\mathcal{O}$. The 
  minus sign differs from the usual picture in the literature, but
  emerges from an improved correspondence formula
  \cite{Mueck02a}. Moreover, our energy momentum
  tensor differs by a minus sign from the convention of BFS.}
\begin{align}
\notag 
  \left\langle T \right\rangle - (\Delta-d)\hat{\phi} 
  \left\langle \mathcal{O} \right\rangle &=
  \left[ \frac{2\hat{g}_{ij}}{\sqrt{\hat{g}}} 
  \frac{\delta}{\delta \hat{g}_{ij}} +
  (\Delta-d) \hat{\phi} \frac1{\sqrt{\hat{g}}} \frac{\delta}{\delta
  \hat{\phi}} \right] \Sren\\
\notag &= \lim_{\rho\to\infty} \left\{ \e{d\rho/l} \left[ 
  \frac{2 g_{ij}}{\sqrt{g}} \frac{\delta}{\delta g_{ij}} +
  (\Delta-d) \phi \frac1{\sqrt{g}} \frac{\delta}{\delta \phi} \right]
  \Gamma \right\} \\
\label{Anom}
  &= \lim_{\rho\to\infty} \left( \e{d\rho/l} \Anom \right) 
  = \hat{\Anom}~.
\end{align}
In passing from the first to the second line we have used the fact
that the integrated anomaly is conformally invariant.

\subsection{Special Case $\lambda=0$}
\label{sp}
There is one logarithmic divergence that has not been addressed so
far and which occurs in the case $\lambda=0$. In
this case, no breakdown of the level 0 descent equation occurs,
although the boundary integral of $\phi^2$ is finite. Hence, this
divergence must be derived in a different fashion. This was considered
also in \cite{Kalkkinen01b}, but the argument there is not complete.

In the case $\lambda=0$, the asymptotic behaviour of the scalar field
is not given by the generic expression \eqref{intro:asympt}, but
follows 
\begin{equation}
\label{sp_asympt}
  \phi(x,r) = \e{-dr/(2l)} \left[ \frac{r}l \hat{\phi}(x)+
  \check{\phi}(x) + \cdots \right]~.
\end{equation}
Thus,
\begin{equation}
\label{sp_dphi}
  \partial_r \phi = -\frac{d}{2l} \phi +\frac1l \e{-dr/(2l)} \hat{\phi}
  +\cdots~,
\end{equation}
but we have also 
\begin{equation}
\label{sp_dphi2}
  \partial_r \phi = \pi = U' + \pi_\Gamma + \cdots~.
\end{equation}
Comparing eqns.~\eqref{sp_dphi} and \eqref{sp_dphi2} and using eqns.\
\eqref{U} and \eqref{U2}, we can read off 
\begin{equation}
\label{sp_pi}
   \pi_\Gamma = \frac1l \e{-dr/(2l)} \hat{\phi} +\cdots~,
\end{equation}
where the ellipses stand for terms that are exponentially suppressed
compared to the one written. In order to find the logarithmic
divergence in terms of the fields living at the cut-off, write
\begin{equation}
\label{sp_piGamma}
  \frac1{\sqrt{g}} \frac{\delta\Gamma}{\delta \phi} = 
  \pi_\Gamma \sim \frac1\rho \phi~,
\end{equation}
from which follows that 
\begin{equation}
\label{sp_Gamma}
  \Gamma \sim \frac1{2\rho} \int d^dx \sqrt{g}\, \phi^2~,
\end{equation}
in addition to any logarithmically divergent terms discussed in Sec.\
\ref{logardiv}, which might arise for even $d$. 
Hence, in the case $\lambda=0$, the renormalized on-shell action is
defined as 
\begin{equation}
\label{sp_Sren}
  \Sren = \lim_{\rho\to\infty} \left( \Gamma + \rho \int d^dx \sqrt{g}\,
  \Hrest - \frac1{2\rho} \int d^dx \sqrt{g}\, \phi^2 \right)~.
\end{equation}

The other special feature of this case is the calculation of the
anomaly. As in Sec.~\ref{logardiv}, we can use the fact that, by
solving the descent equations, the hamiltonian constraint has been
reduced to eqn.~\eqref{anom_eq}, but $\Ham_\Gamma$ now has the form
\begin{equation}
\label{sp_Hgamma}
  \Ham_\Gamma = 8 q^{ij}_\Gamma q_{[0]ij} -\frac8{d-1} q_\Gamma
  q_{[0]} + \pi_\Gamma \pi_{[0]} + \frac12 \pi_\Gamma^2 + \cdots~.
\end{equation}
We have to include the term $\pi_\Gamma^2$, because in the power
counting it has the same strength as the other terms. This yields 
\begin{align}
\notag 
  \left\langle T \right\rangle +\frac{d}2 \hat{\phi} 
  \left\langle \mathcal{O} \right\rangle &=
  \left[ \frac{2\hat{g}_{ij}}{\sqrt{\hat{g}}} 
  \frac{\delta}{\delta \hat{g}_{ij}} 
  -\frac{d}2 \hat{\phi} \frac1{\sqrt{\hat{g}}} \frac{\delta}{\delta
  \hat{\phi}} \right] \Sren\\
\notag &= \lim_{\rho\to\infty} \left\{ \e{d\rho/l} \left[ 
  \frac{2 g_{ij}}{\sqrt{g}} \frac{\delta}{\delta g_{ij}} 
  -\frac{d}2 \phi \frac1{\sqrt{g}} \frac{\delta}{\delta \phi} \right]
  \Gamma \right\} \\
\label{sp_Anom}
  &= \lim_{\rho\to\infty} \left[ \e{d\rho/l} \left(\Anom -\frac{l}2
  \pi_\Gamma^2 \right) \right] 
  = \hat{\Anom} -\frac1{2l} \hp^2~.
\end{align}
As before, we have used the conformal invariance of all logarithmic
counterterms. In eqn.~\eqref{sp_Anom}, $\Anom$ is defined by
eqn.~\eqref{A} [eqn.~\eqref{Adef} does not hold here], so that 
$\hat{\Anom}$ denotes the gravitational conformal anomaly for even
$d$, and the $\hp^2$ term is the matter conformal anomaly.

\section{One-point Functions}
\label{one_pt}
\subsection{Exact One-point Functions}
\label{exact_1pt}
One of the main results of the standard method of holographic
renormalization is that it yields formal expressions for \emph{exact}
one-point functions \cite{Bianchi01b}. Exact one-point functions
depend on finite sources and thus contain information about all higher
point functions 
of the theory. It is therefore desirable to obtain the same results
using the Hamilton-Jacobi approach. This is straightforward, given
the renormalized on-shell action, eqn.~\eqref{S_ren} [or 
eqn.~\eqref{sp_Sren} in the case $\lambda=0$], and we shall demonstrate the
procedure by considering the scalar one-point function for the cases
$d=4$, $\lambda=1$ and $\lambda=0$, $d$ arbitrary.

Generally, we obtain
\begin{align}
\notag 
  \left\langle \mathcal{O}(x) \right\rangle &= -\frac1{\sqrt{\hat{g}}}
  \frac{\delta S_{\text{ren}}}{\delta \hat{\phi}(x)} \\
\notag 
  &= - \lim_{\rho\to\infty} \left[ \e{(d/2+\lambda)\rho/l}
  \frac1{\sqrt{g}} \frac{\delta}{\delta \phi(x,\rho)} \
  \left( \Gamma +\rho\int d^dx \sqrt{g}\, \Hrest \right)\right]\\
\label{one_pt:O}
  &= - \lim_{\rho\to\infty} \left[ \e{(d/2+\lambda)\rho/l} 
  \left( \pi_\Gamma + \rho \frac{\delta\Hrest}{\delta \phi} \right)
  \right]~,
\end{align}
but, in practise, it is more useful to substitute $\pi_\Gamma$ from
the split form \eqref{split_pi} of the hamilton equation of motion
\eqref{momenta_pi}, 
\begin{equation}
\label{one_pt:pi_Gamma}
  \pi_\Gamma = \partial_r \phi - \pi_{[0]} - \cdots - \pi_{[2n]}~,
\end{equation}
because at this stage $\pi_\Gamma$ is not known explicitly in terms of
the source. Moreover, it is necessary to write down an asymptotic
expansion for $\phi$ beyond leading order, namely of the form of eqn.\
\eqref{intro:asympt}. However, in contrast to the standard method, we
do not have to solve for the subleading coefficients using the
equations of motion, an ansatz with the subleading 
$f_1$, $f_2$, etc.\ undetermined is sufficient. In practise, finding a
general ansatz asks for some care, because generic bulk interactions might
generate all sorts of exponents in the subleading terms. However, for
our purposes, we shall regard this step as done.

Specializing to $d=4$ and $\lambda=1$, the potential $V$ has the form
\begin{equation}
\label{one_pt:V}
  V(\phi) = - \frac3{l^2} - \frac3{2l^2} \phi^2 + \frac{v_3}{6} \phi^3
  + \frac{v_4}{4!} \phi^4 +\cdots~.
\end{equation}
This example is a generalization of the potential generating the GPPZ flow
\cite{Girardello99a}. According to the results of Sec.~\eqref{powdiv},
the function $U(\phi)$ occuring in the leading counterterm has the form
\begin{equation}
\label{one_pt:U}
  U(\phi) = -\frac3{2l} -\frac1{2l} \phi^2 +\frac{lv_3}{6} \phi^3
  +\frac{u_4}{4!} \phi^4 +\cdots~.
\end{equation}
Remember that the coefficient $u_4$ remains undetermined. Similarly,
we have for the functions $\Phi(\phi)$ and $M(\phi)$,
\begin{align}
\label{one_pt:Phi} 
  \Phi(\phi) &= -\frac{l}8 +\frac12 \Phi_2 \phi^2 +\cdots~,\\
\label{one_pt:M}
  M(\phi) &= m_0 +\cdots~,
\end{align}
with $\Phi_2$ and $m_0$ undetermined. The relevant asymptotic expansion
for $\phi$ is of the form 
\begin{equation}
\label{one_pt:asympt}
  \phi(x,r) = \e{-r/l} \hp + \e{-2r/l} \phi_1 + \e{-3r/l} \phi_2 +
  \frac{r}l \e{-3r/l} \psi_2 + \cdots~,
\end{equation}
while it is sufficient to use the leading term for the metric, 
\begin{equation}
\label{one_pt:g}
  g_{ij}(x,r) = \e{2r/l} \hat{g}_{ij} + \cdots~.
\end{equation}

The unresolved remainder from the descent equations is given by 
eqns.~\eqref{Hrest_4}, \eqref{Hrest2} and \eqref{Hrest4},
\begin{equation}
\label{one_pt:Hrest}
  \Hrest = -\frac{l^2}{16} \left( \frac13 R^2 -R_{ij} R^{ij} \right)
  -\frac12 \nabla^i \phi \nabla_i \phi - \frac1{12} R\phi^2 +
  \left( \frac{l^2 v_3^2}8 -\frac1{3l^2} -\frac{v_4}{24} \right) \phi^4~.
\end{equation} 
Thus, after substituting everything into eqn.~\eqref{one_pt:O}, one finds 
\begin{equation}
\label{one_pt:Oexpl}
\begin{split}
  \left\langle \mathcal{O} \right\rangle &= \lim_{\rho\to\infty}
  \left\{ \e{\rho/l} \left(
  \frac1l \phi_1 +\frac{lv_3}2 \hp^2 \right) + \rho
  \left[ \frac2{l^2} \psi_2 - \hat{\nabla}^2 \hp + \frac{\hat{R}}6 \hp 
  + \left( \frac{v_4}6 +\frac{4}{3l^2} - \frac{l^2v_3^2}2 
  \right) \hp^3 \right] \right. \\
  &\quad \left.+ \frac2l \phi_2 -\frac1l \psi_2 + lv_3 \hp\phi_1
  +\frac{u_4}{6} \hp^3 -m_0 \hat{\nabla}^2 \hp +\Phi_2 \hat{R} \hp \right\}~.
\end{split}
\end{equation}
Notice that the leading divergent term, proportional to $\e{2\rho/l}$,
has explicitly cancelled. Moreover, since we have removed all
divergences from the on-shell action (as was proven in
Sec.~\ref{logardiv}), the correlation functions must be finite.  
Hence, the remaining two divergent terms in eqn.\ \eqref{one_pt:Oexpl} 
must vanish by construction. Thus, we obtain the
coefficients $\phi_1$ and $\psi_2$ of the leading asymptotic solution, 
\begin{equation}
\label{one_pt:coeffs}
  \phi_1= -\frac{l^2v_3}2 \hp^2~, \qquad \psi_2 = \frac{l^2}2 \left[
  \hat{\nabla}^2 \hp -\frac{\hat{R}}6 \hp - 
  \left( \frac{v_4}6 +\frac{4}{3l^2} - \frac{l^2v_3^2}2 
  \right) \hp^3\right]~.
\end{equation}
Substituting eqn.~\eqref{one_pt:coeffs} into eqn.\
\eqref{one_pt:Oexpl}, we finally obtain
\begin{equation}
\label{one_pt:Ofin} 
  \left\langle \mathcal{O} \right\rangle = \frac2l \phi_2 - 
  \left( m_0+\frac{l}2 \right) \hat{\nabla}^2 \hp 
  + \left(\Phi_2 + \frac{l}{12} \right) \hat{R} \hp + \left(
  \frac{u_4}{6} +\frac {lv_4}{12} +\frac2{3l} -\frac{3l^2v_3^2}4
  \right) \hp^3~,
\end{equation}
where the non-local behaviour is encoded in the undetermined
coefficient $\phi_2$ of the subleading series solution, which is
obtained by imposing regularity of the 
field in the bulk. The arbitrary parameters
$u_4$, $\Phi_2$ and $m_0$ represent the scheme dependence. In
particular, there exists a renormalization scheme, in which $\phi_2$
alone represents the exact scalar one-point function.  

This example underlines the reverse approach of the HJ method with
respect to the standard method of holographic renormalization. 
While, in the latter, one first determines the sub-leading coefficients
in the asymptotic expansion of the fields from the equations of
motion and then proceeds to renormalize the
on-shell action, in the former one first renormalizes the on-shell
action and then obtains the sub-leading coefficients from the
finiteness of the correlation functions. One could ask whether the
subleading coefficients obtained in eqn.\ \eqref{one_pt:coeffs} agree
with those calculated in the standard method, \emph{i.e.}, whether
they are consistent with the equations of motion. In the present
example they obviously do, as seen by comparison with
\cite{Bianchi01b}, but the answer is affirmative in general, because
the coefficients follow uniquely by the procedure given above.

The case $\lambda=0$ is even simpler. Since the
renormalized on-shell action is given by eqn.~\eqref{sp_Sren}, we
obtain
\begin{align}
\notag
  \left\langle \mathcal{O} \right\rangle &= 
  -\frac1{\sqrt{\hat{g}}} \frac{\delta \Sren}{\delta \hat{\phi}} \\
\notag 
  &= - \lim_{\rho\to\infty} \left[ \e{d\rho/(2l)} \frac{\rho}l 
  \frac1{\sqrt{g}} \frac{\delta}{\delta \phi} \left( \Gamma - 
  \frac1{2\rho} \int d^dx \sqrt{g}\, \phi^2 \right) \right] \\
\label{sp_1pt}
  &= - \lim_{\rho\to\infty} \left[ \e{d\rho/(2l)} \frac{\rho}l 
  \left(\pi_\Gamma - \frac1\rho \phi \right)  \right] =\frac1l \check{\phi}~.
\end{align}
In the second line we have used the fact that $\Hrest$ does not depend
on $\phi$ in the present case,
while the last result follows from eqns.~\eqref{sp_asympt} 
and \eqref{sp_pi}. This result is in agreement with
our expectation, \emph{viz.}, $\check{\phi}$
corresponds to the vev in the boundary field theory. 

The same procedure can be used to calculate the exact one-point
function for the energy momentum tensor, $\langle T_{ij} \rangle$. The
general formula is (for $\lambda\neq0$)
\begin{align}
\notag
  \left\langle T_{ij} \right\rangle &= -\frac2{\sqrt{\hat{g}}}
  \frac{\delta S_{\text{ren}}}{\delta \hat{g}^{ij}} \\
\notag 
  &= \lim_{\rho\to\infty} \left[ \e{(d-2)\rho/l} \left( -\frac2{\sqrt{g}}
  \frac{\delta}{\delta g^{ij}} \right) \left( \Gamma +\rho \int d^dx
  \sqrt{g}\, \Hrest \right) \right]\\
\label{one_pt:T}
  &=  \lim_{\rho\to\infty} \left[ \e{(d-2)\rho/l} \left( 2
  q_{\Gamma ij} - \frac{2\rho}{\sqrt{g}}
  \frac{\delta}{\delta g^{ij}} \int d^dx \sqrt{g}\, \Hrest \right)
  \right]~.
\end{align}

For $d=4$, the full analysis is
quite tiresome and involves also scheme dependent counterterms, which are
quadratic in the boundary curvature. We shall not perform the explicit
expansion, since, for the GPPZ and the Coulomb branch
flows, the result is known \cite{Bianchi01b}. We only point
out that the trace of the energy momentum tensor one-point function is
given by the anomaly formulae, eqns.~\eqref{Anom} and
\eqref{sp_Anom}, and its divergence satisfies the Ward identity
\begin{equation}
\label{Ward}
  \hat{\nabla}_j \langle T^{ij} \rangle + \langle \mathcal{O} \rangle
  \hat{\nabla}^i \hp = 0~.
\end{equation}
Eqn.~\eqref{Ward} 
follows from the constraint \eqref{con_Hi} and the fact that 
all counterterms are diffeomorphism invariant, because they are
expressed as covariant boundary integrals.

\subsection{One-point Functions in Holographic RG Flows}
\label{RG_1pt}

The calculation of one-point functions in holographic RG flows, \emph{i.e.}
those with the sources set to their background values, 
is particularly simple. Holographic RG flows are solutions of
the bulk field equations of the form 
\begin{equation}
\label{RGflow}
  \phi= \phi(r)~,\qquad g_{ij} = \e{2A(r)} \eta_{ij}~,
\end{equation}
which are obtained by solving the first order equations
\cite{Freedman99a,DeWolfe00b,Skenderis99} 
\begin{equation}
\label{RGflow_eq}
  \partial_r \phi= W'(\phi)~,\qquad \partial_r A = -\frac2{d-1} W(\phi)~,
\end{equation}
if the potential $V$ can be written in terms of a function $W(\phi)$ as
\begin{equation}
\label{VW}
  V(\phi) = -\frac{d}{d-1} W^2 +\frac12 \left(W'\right)^2~.
\end{equation}
In addition, for $r\to\infty$, the solution should approach a fixed
point of $W$ with negative value, so that the bulk becomes
asymptotically AdS. 

It is obvious that eqn.~\eqref{VW} is identical
to the level zero constraint, eqn.~\eqref{H0}, when $W$ is
substituted for $U$. Similarly, eqn.~\eqref{RGflow_eq} are the
Hamilton equations of motion, \eqref{momenta_q} and
\eqref{momenta_pi}. This means that $W$ necessarily is a solution of
the level zero descent equation, but in order to be used in the
leading counterterm, it must have an expansion 
\begin{equation}
\label{Wexp1}
  W(\phi) = -\frac{d-1}{2l} +\frac1{2l} \left( -\frac{d}2 +\lambda
  \right) \phi^2 +\cdots~.
\end{equation}
In particular, the quadratic coefficient must be
$-d/2+\lambda$ with $\lambda>0$.\footnote{This is the only detail,
  which is not fixed by the level zero constraint, the asymptotically
  AdS metric and the existence of the fixed point.}
Let us assume for the moment that this is the case. Then, after identifying
$W\equiv U$, the calculation of the scalar one-point function becomes
trivial: 
\begin{equation}
\label{RG_1pt_O1}
  \left\langle \mathcal{O} \right\rangle =  
  - \lim_{\rho\to\infty} \left[ \e{(d/2+\lambda)\rho/l}
  \left(\partial_r \phi - U'  + \rho \frac{\delta \Hrest}{\delta \phi}
  \right)\right]=0~.
\end{equation} 
The second equality holds because of eqn.~\eqref{RGflow_eq} and because
neither the higher order counterterms nor $\Hrest$ contribute in the
Poincar\'e invariant background. 

Similarly, for the energy momentum tensor one finds
\begin{equation}
\label{RG_1pt_T1}
  \left\langle T_{ij} \right\rangle = \lim_{\rho\to\infty} \left[
  \e{(d-2)\rho/l} \left( -\frac{d-1}2 \partial_r A -U \right) g_{ij}  
  \right] =0~. 
\end{equation}
Eqn.~\eqref{RG_1pt_O1} explicitly shows that prepotentials $W$ with the
expansion \eqref{Wexp1} are flows generated by the addition of a
relevant operator to a CFT Lagrangian.

Notice also that
the choice $U=W$ for the leading counterterm means that we fix the
coefficients in $U$, which are left undetermined by a possible
breakdown of the level zero descent equation. 
This corresponds to a \emph{supersymmetric} renormalization scheme,
which is supported by the vanishing of the energy momentum tensor 
\eqref{RG_1pt_T1} and
of $S_{\text{ren}}$ for a bulk solution of the form
\eqref{RGflow_eq}. 

The second possibility, which we shall now discuss, is that $W$ has the
expansion 
\begin{equation}
\label{Wexp2}
  W(\phi) = -\frac{d-1}{2l} +\frac1{2l} \left( -\frac{d}2 -\lambda
  \right) \phi^2 +\cdots~.
\end{equation}
Here, $\lambda$ is again positive. 
In this case, the background scalar field has the asymptotic behaviour 
\begin{equation}
\label{asympt2} 
  \phi(r) = \e{-(d/2+\lambda)\rho/l} \cp +\cdots~,
\end{equation}
which is entirely sub-leading with respect to the generic behaviour. 
Then, eqn.~\eqref{one_pt:O} yields 
\begin{align}
\notag
  \left\langle \mathcal{O}(x) \right\rangle &=  
  - \lim_{\rho\to\infty} \left[ \e{(d/2+\lambda)\rho/l}
  \left(W' - U'  + \rho \frac{\delta \Hrest}{\delta \phi}
  \right)\right]\\ 
\label{RG_1ptO2}
  &= - \lim_{\rho\to\infty} \left[ \e{(d/2+\lambda)\rho/l} \left(
  -\frac{2\lambda}l \phi +O(\phi^2) \right) \right] 
  = \frac{2\lambda}l \cp~. 
\end{align}
Notice that the terms in $\Hrest$, which do not contain the boundary
curvature, stem from a breakdown of the level zero equation and involve
$\phi$ with at least cubic power, and these terms vanish in
the $\rho\to \infty$ limit due to the weak asymptotic behaviour of
$\phi$. Eqn.~\eqref{RG_1ptO2} explicitly demonstrates that holographic
RG flows with $W$ of the form \eqref{Wexp2} correspond to deformations
of the boundary CFT by switching on a vev of the scalar operator. 
Also in this case the one-point function of the
energy momentum tensor vanishes. From eqn.~\eqref{one_pt:T} one finds
\begin{align}
\notag
  \left\langle T_{ij} \right\rangle &= \lim_{\rho\to\infty} \left\{ 
  \e{(d-2)\rho/l} \left[ (W-U) g_{ij} + 2\rho g_{ij} O(\phi^3) \right]
  \right\} \\
\label{RG_1pt_T2}
  &= \lim_{\rho\to\infty} \left\{ 
  \e{d\rho/l} \left[ -\frac{\lambda}l \phi^2 \hat{g}_{ij} + 2\rho
  \hat{g}_{ij} O(\phi^3) \right] \right\} = 0~,
\end{align}
again due to the weak asymptotic behaviour of $\phi$. 

Last, let us comment briefly on the case, where $W$ has the expansion
\eqref{Wexp1} with $\lambda=0$. The asymptotic behaviour of the RG
flow solution is given by eqn.~\eqref{sp_asympt} with
$\hp=0$. The scalar one-point function has been calculated in 
Sec.~\ref{exact_1pt} and is given by eqn.~\eqref{sp_1pt}. The energy
momentum tensor again vanishes, since, in the additional counterterm,
the scalar field is not strong enough to compensate the $1/\rho$ factor.

\section{Contributions of the Vector Sector}
\label{vectors}
In the following we will add the vector sector of the theory
to the gravity and scalar sectors discussed so far. We will show how
the Ward identities previously derived are modified by the presence of
interacting vector fields, including expected new contributions to the
conformal anomaly, and discuss the additional current Ward identity
associated with the $U(1)$ gauge invariance. In this way we
provide an alternative derivation of the holographic chiral anomaly,
already obtained in \cite{Bianchi01b} by using the standard approach
to holographic renormalization and in \cite{KOW} by explicitly
analyzing some dual supergravity solutions.

The physics of an RG flow in the presence
of a residual $U(1)$ $R$-symmetry is encoded in a bulk $d+1$ dimensional
action, which contains a ``massive'' vector field $A_\mu$ together with a
St\"uckelberg field $\alpha$, interacting with scalars and gravity (see
\cite{anatomy,Bianchi01b}). Thus, our
starting point is the gauge-invariant
action \eqref{app:action}. The Hamiltonian constraints
associated with this are derived and summarized in the
Appendix. Before we begin the analysis, let us say two words on the
issue of the asymptotic scaling of the fields. When we include the new
fields, the equations of motion become rather intricately coupled,
and we have to make sure that interactions between the fields do not
influence their leading asymptotic behaviour. 
As a simplification, one can consider gravity
and vectors in some fixed scalar background. This is the assumption
made by BFS, and we borrow their analysis of the leading behaviour of
the vector sector fields. From \cite{Bianchi01b}, we see that
\begin{align}
\label{B_asym}
  B_i (x,r) &= \hat{B}_i (x)+\cdots\\
\label{alpha_asym_GPPZ}
  \alpha (x,r) &= \hat{\alpha}(x) + \cdots &&\mathrm{(GPPZ)}\\
\label{alpha_asym_CB}
  \alpha (x,r) &=\frac{r}{l} \hat{\alpha}(x) + \check{\alpha} (x) + \cdots
  &&\mathrm{(CB)}~.
\end{align}
for $r\to\infty$.

\subsection{Descent Equations and Conformal Anomaly}
\label{vect_conf_anom}
As discussed in Sec.~\ref{regren}, in order to analyze the descent equations
obtained from the constraint $\Ham=0$, it is convenient to group the terms of
the local part of the on-shell action into different levels. The
lowest possible counterterms of the vector sector are of level two and
read
\begin{equation}
  S_{[2]}^v  =  \int d^dx \sqrt{g} \left[ \frac{1}{2}N(\phi)g^{ij}B_iB_j+
  \frac{1}{2}P(\phi)g^{ij}\nabla_i\phi B_j\right]~.
\label{S2v}
\end{equation}

After computing the momenta stemming from the sum of \eqref{S0}, \eqref{S2},
and \eqref{S2v}, and inserting them into eqn.~\eqref{con_H},
one can see that the level zero equation is unchanged, and one can
always solve the level two contributions by setting $P(\phi)=0$. This
is justified, since we do not expect counterterms linear in $B_i$. In
this way the analysis of the gravity-scalar sector remains unchanged with
respect to Sec.~\ref{powdiv}. The new equation to be solved is
\begin{equation}
{\Ham}_{[2]}^v = \left( \frac{U'N'}{2}-\frac{d-2}{d-1}UN+\frac{N^2}{2K}-
               \frac{\M^2}{2}\right)\, B^i B_i=0~ .
\end{equation}
We proceed in the usual way by further expanding $U,N$ and $\M^2$ in powers of
$\phi$. Notice that, for the flows we are interested in, $K(\phi)=O(1)$
and $\M^2(\phi)= O(\phi^2)$. Using the solution of $U$ obtained in
Sec.~\ref{powdiv}, we first get
\begin{equation}
  N_0 \left(\frac{N_0}{2K_0}+\frac{d-2}{2l}\right) =  0~.
\end{equation}
Again we shall compare to pure AdS/CFT results in order to select one of
the two possibilities in the above equation. Looking for instance in
\cite{Mueck98-2} we learn that one should have $N_0=0$ for
asymptotically massless vector fields. Proceeding up
to quadratic order in $\phi$, we unambiguously determine
\begin{align}
  N_1 & =  0\\
N_2 & =  \frac{l}{\lambda-1}\M^2_2~.
\end{align}
A breakdown occurs, if $\lambda=1$, thus giving the contribution
\begin{equation}
  \Hrest  =  \frac{1}{2}\M^2_2\phi^2B^i B_i +
  O(\phi^3,B^4) \qquad (\lambda=1)~.
\label{Banomaly}
\end{equation}
Following the general analysis of Sec.~\ref{logardiv} this yields a
logarithmic divergence of $\Gamma$ and a contribution to the conformal
anomaly for scalar operators with $\lambda=1$.

Proceeding to the next levels increases the number of invariants very
quickly, and a complete analysis already at level four would be
extremely tedious. The gravity-scalar sector has been analysed in
\cite{Kalkkinen01b}, and in the following we restrict to the vector sector
in a fixed scalar background. A possible basis of independent
invariants for the level four on-shell action is the following
\begin{equation}
\begin{split}
  & (B_iB^i)^2,\nabla^i B_iB_jB^j,B_iB_j\nabla^i B^j,F_{ij}F^{ij},
  (\nabla^i B_i)^2,\\
  & \nabla_i B_j \nabla^j B^i,
  \nabla_i B_j \nabla^i B^j,R_{ij}B^iB^j,\nabla^iB^jR_{ij}~.
\end{split}
\label{level4inv}
\end{equation}
It turns out that the resulting level four descent equations, which we
do not write here, can be consistently solved
setting to zero the coefficients of all the terms except for
$F_{ij}F^{ij}$. This leads to the descent equation
\begin{equation}
\label{F2}
  \Ham^{F}_{[4]} = \left( \frac{d-4}{4l}G_0 - \frac{1}{4}K_0
  \right) F_{ij} F^{ij} = 0~,
\end{equation}
where $G_0$ is the constant part of the coefficient of the
$F_{ij}F^{ij}$ counterterm. The descent equation \eqref{F2} breaks for
$d=4$, leading to a logarithmic divergence of $\Gamma$ and to a
contribution to the conformal anomaly.

Hence, we have found the following anomaly contributions from the
vector sector, up to level four:
\begin{equation}
\label{vec_anom}
\hat\Anom =
\begin{cases}
  -\frac{1}{2}l \M^2_2\hat{\phi}^2\hat{B}_i \hat{B}^i & (\lambda=1)~, \\
  \frac{1}{4}l K_0\hat{F}_{ij} \hat{F}^{ij} & (d=4)~.
\end{cases}
\end{equation}
They agree with the logarithmic counterterms of \cite{Bianchi01b}, and
the $d=4$ contribution was determined also in \cite{marika,Kalkkinen00}.

\subsection{Conformal Ward Identity}
\label{vect_conf_ward}
Let us show how the anomalous conformal Ward identity is affected by the
presence of currents. First, notice that the level zero terms of the
$E^i$ and $\pi_\alpha$ momenta vanish, because the vector part of the
on-shell action starts at level two [cf.~eqn.~\eqref{S2v}].
Therefore, ${\Ham}_\Gamma$ is the same at leading order as in
eqn.~\eqref{Hgamma}. This implies that exactly the same form of conformal
Ward identity holds here, though with the additional anomaly
contributions derived in Sec.~\ref{vect_conf_anom}, 
eqn.~\eqref{vec_anom}, namely
\begin{equation}
  \langle T\rangle - (\Delta-d)\,\hat{\phi}\,\langle
  \mathcal{O}\rangle = \hat{{\Anom}}~.
\label{Anom2}
\end{equation}
This might seem slightly odd. However, it is \emph{consistent} with
the scaling dimensions assigned to the various fields, which are read
as usual from the leading $r$ dependence of the asymptotic
expansions. In particular, recall that we have the following
behaviours near the boundary ($r\to\infty$)
\begin{equation}
  \phi \sim \hat{\phi}\,\e{-(d-\Delta)r/l} \qquad\quad
  A_i\sim\hat{A}_i \qquad\quad \alpha \sim \hat{\alpha}~.
\end{equation}
This implies that the sources $\hat{A}_i$ and $\hat{\alpha}$ do not
transform under Weyl rescaling, and eqn.~\eqref{Anom2} correctly
reflects this fact. On the other hand, it is well known that a
conserved current should have conformal dimension
$\tilde{\Delta}=d-1$, and thus couple to a vector source of weight
one. In the context of the AdS/CFT  correspondence this is realized by
using as source the (rescaled) vector field in a local Lorentz frame,
$A_a = e_{a}^{\;i} A_i$ \cite{Mueck98-2}. Hence, we should have used
the frame fields $A_a$ and the vielbeins $e^{a}_{\;i}$ as
``coordinates'' in the Hamiltonian treatment of the system. This would
complicate the analysis unnecessarily, but it is easy to correct 
eqn.~\eqref{Anom2} for this misuse. In fact, setting $\Sren[g_{ij},B_i] =
\tilde{S}_{\text{ren}}[e^{a}_{\;i},B_a]$, a straightforward
application of the chain rule gives
\begin{equation}
  \langle \tilde{T}^{a}_{\;i} \rangle  =  T^{a}_{\;i} +  \hat{B}_i
  \langle J^{a} \rangle~.
\end{equation}
After tracing the above expression and inserting it into 
eqn.~\eqref{Anom2}, we get
\begin{equation}
\langle \tilde{T}\rangle - (\Delta-d)\,\hat{\phi}\,\langle \mathcal{O}\rangle
-\hat{B}_{a}\langle J^{a}\rangle   =
 \hat{{\Anom}}~.
\label{totaltrace1}
\end{equation}
Anticipating the next section,  we can use gauge invariance
to rewrite eqn.~\eqref{totaltrace1} in its standard field theory form
as
\begin{equation}
  \langle \tilde{T}\rangle - (\Delta-d)\,\hat{\phi}\,\langle \mathcal
 {O}\rangle -\hat\alpha\langle \mathcal{O}_\alpha \rangle
 -\hat{A}_{a}\langle J^{a}\rangle = \hat{{\Anom}}~,
\end{equation}
where we have used the
freedom to add a total derivative to the anomaly.

\subsection{Diffeomorphism and Gauge Ward Identities}
\label{vect_diff_gauge_ward}
In this section we shall derive the Ward identities that follow from
the hamiltonian constraints associated with bulk diffeomorphisms
and gauge invariance. From the constraints \eqref{con_G} and
\eqref{con_Hi} and the premises on the power-divergent counterterms
stated in Sec.~\ref{powdiv} we get
\begin{align}
  \nabla_i E^i_\Gamma+ \pi_{\alpha\,\Gamma} & = 0
  \label{gauge_gamma}\\
  2 \nabla_j q^j_{i\,\Gamma} -\nabla_i\phi\pi_\Gamma-F_{ij}E^j_\Gamma-
  B_i \pi_{\alpha\,\Gamma}  & =  0~,
\label{diffeos_gamma}
\end{align}
respectively.
How these relations translate into Ward identities for the finite field theory
quantities depends, as usual, on whether logarithmically divergent terms
hidden in $\Gamma$ possibly violate the constraints. Let us
assume here that all logarithmic counterterms are obtained by the breakdown
of the descent equations as described in Sec.~\ref{vect_conf_anom}. These
counterterms are, by construction, both covariant and gauge invariant. Thus,
the relations \eqref{gauge_gamma} and \eqref{diffeos_gamma} directly
translate into the following Ward identities for gauge and
diffeomorphism invariance, respectively,
\begin{align}
\label{gauge_ward}
  \hat{\nabla}_i\langle J^i\rangle
  +\langle \mathcal {O}_\mathrm{\alpha}\rangle &= 0~,\\
\label{diffeo_ward}
  \hat{\nabla}^j\langle T_{ij}\rangle
  +\langle \mathcal {O}\rangle\nabla_i\hat{\phi}
  +\langle J^j\rangle \hat{F}_{ij}
  +\langle \mathcal {O}_\mathrm{\alpha}\rangle (\hat{A}_i +\partial_i
  \hat{\alpha}) &=0~.
\end{align}
The expression given by BFS as the diffeomorphism Ward identity 
(eqn.~(4.19) in \cite{Bianchi01b}) is obtained by substituting
eqn.~\eqref{gauge_ward} into \eqref{diffeo_ward}.

\subsection{Special Case}
\label{sp_CB}

As in Sec.~\eqref{sp}, one has to be more careful, if the leading
term of a field is logarithmic, as is the case for $\alpha$ in the Coulomb
branch flow with a \emph{fixed scalar background} [cf.~eqn.\
\eqref{alpha_asym_CB}]. In fact the following analysis
applies only if we consider gravity and vectors in a fixed
scalar background (denoted with a bar).
That is, we are neglecting the back reaction of a possible scalar
source turned on. This would complicate the situation considerably, as
can be seen by inserting the leading behaviour of $\alpha$ from eqn.\
\eqref{alpha_asym_CB} into the action: The source $\hat{\alpha}$ contributes
to the asymptotic mass of the scalar $\phi$ violating the assumption
that the leading order behaviour does not depend on interactions. 

Let us discuss this special case and derive
the anomalous Ward identities for the CB flow.
Using eqns.~\eqref{alpha_asym_CB} and \eqref{momenta_a} and following
similar steps as in Sec.~\ref{sp}, we find
\begin{equation}
\label{pi_alpha_CB}
  \frac{1}{\sqrt{g}} \frac{\delta \Gamma}{\delta \alpha} =
  \pi_{\alpha\,\Gamma} =\frac1l \M^2 \hat{\alpha} \sim
  \frac{1}{\rho} \M^2 \alpha~.
\end{equation}
From this follows that
\begin{equation}
  \Gamma \sim \frac{\M_2^2}{4\rho}\int d^dx \sqrt{g}\, \bar{\phi}^2\,\alpha^2~,
\end{equation}
in addition to the logarithmically divergent terms stemming from $\Hrest$.
The renormalized on-shell action is given by
\begin{equation}
\label{sren_CB}
  \Sren = \lim_{\rho\to \infty} \left( \Gamma +\rho \int d^d x \sqrt {g}\Hrest
  -\frac{\M_2^2}{4\rho}\int d^dx \sqrt{g}\,\bar{\phi}^2\,\alpha^2\right)~.
\end{equation}
Notice that, with the scalar background fixed, the additional
counterterm in eqn.~\eqref{sp_Sren}
is irrelevant, and we have not included it here.

The additional counterterm in eqn.~\eqref{sren_CB} is not gauge
invariant, which will violate the Ward identity \eqref{gauge_ward}. In
fact, it follows straightforwardly from eqns.~\eqref{gauge_gamma} and
\eqref{pi_alpha_CB} that
\begin{equation}
  - \e{-d\rho/l} \hat{\nabla}_i \langle J^i \rangle + \frac1{2l}\M_2^2
    \bar{\phi}^2 \hat{\alpha}=0~.
\end{equation}
Hence, we obtain
\begin{equation}
\label{chiral_anom}
   \hat{\nabla}_i \langle J^i \rangle = \frac1{2l}\M_2^2
   \cp^2 \hat{\alpha} = \frac1l \hat{\alpha}~,
\end{equation}
where for the last equality we have substituted the specific values
$\M_2^2=12$, $\cp=-1/\sqrt{6}$ for the CB flow \cite{Bianchi01b}.

As an aside, we can also compute the vev of the operator dual to
$\alpha$ for the CB flow,
\begin{align}
\notag
   \langle \mathcal{O}_\alpha \rangle &=
   - \frac{1}{\sqrt{\hat{g}}} \frac{\delta \Sren}{\delta \hat{\alpha}} \\
\notag
   &= - \lim_{\rho\to\infty} \left[ \e{d\rho/l}
    \frac{\rho}{l}\frac1{\sqrt{g}} \frac{\delta}{\delta \alpha}
    \left( \Gamma - \frac{\M_2^2}{4\rho} \int d^d x \sqrt{g}\,
    \bar{\phi}^2\,\alpha^2\right)\right] \\
\notag
   &= -\lim_{\rho\to\infty} \left[
     \e{d\rho/l} \frac{\rho}{l} \left( \pi_{\alpha\,\Gamma}
     -\frac{\M_2^2}{2\rho}
     \bar{\phi}^2\,\alpha \right)\right]\\
\label{alpha_vev_CB}
   &=\lim_{\rho\to\infty} \left[
     \e{d\rho/l}\frac{\M_2^2}{2l}
     \bar{\phi}^2\,\check{\alpha}\right] = \frac1{2l}\M_2^2 \cp^2
     \check{\alpha} = \frac1l \check{\alpha}~.
\end{align}

The results \eqref{gauge_ward} and \eqref{chiral_anom} were derived in
\cite{Bianchi01b} within the standard approach to holographic
renormalization and represent the holographic realization of the
chiral symmetry breaking. As also noticed in \cite{Brandhuber00a,KOW},
it is clear that the
spontaneous breaking of gauge symmetry in the CB case is essentially
dual to a Higgs mechanism in the bulk, with $\bar{\phi}^2$ providing a mass
for $B_i$ because of its weak asymptotic scaling.

One might worry that the leading logarithmic behaviour will upset the
diffeomorphism Ward identity \eqref{diffeo_ward}, but the following
calculation explicitly confirms it also in this case. From 
eqns.~\eqref{diffeos_gamma}, \eqref{sren_CB} and \eqref{pi_alpha_CB}
follows 
\begin{equation}
\begin{split}
  \e{-d\rho/l} \hat{\nabla}_j T^j_{\;i} + \frac1{4\rho} \M_2^2
  \partial_i (\phi^2 \alpha^2)
  + \e{-d\rho/l} \partial_i \hp \langle \mathcal{O} \rangle -
  \frac1{2\rho} \M_2^2 \phi \partial_i \phi \alpha^2 & \\
  + \e{-d\rho/l}
  \hat{F}_{ij} \langle J^j \rangle - (A_i +\partial_i \alpha) \frac1{2l}
  \M_2^2 \phi^2 \hat{\alpha} &=0~.
\end{split}
\end{equation}
Using the explicit results \eqref{chiral_anom} and
\eqref{alpha_vev_CB} this yields
\begin{multline}
  \e{-d\rho/l} \left[ \hat{\nabla}_j T^j_{\;i} + \hat{\nabla}_i \hp \langle
  \mathcal{O} \rangle + \hat{F}_{ij} \langle J^j \rangle - \hat{A}_i
  \hat{\nabla}_j \langle J^j \rangle \right]
  + \frac1{2l} \M_2^2 \phi^2 \partial_i \alpha \left( \frac{l}{\rho}
  \alpha - \hat{\alpha} \right) \\
  =  \e{-d\rho/l} \left[ \hat{\nabla}_j T^j_{\;i} + \hat{\nabla}_i \hp \langle
  \mathcal{O} \rangle + \hat{F}_{ij} \langle J^j \rangle - \hat{A}_i
  \hat{\nabla}_j \langle J^j \rangle
  + \partial_i \hat{\alpha}\langle \mathcal{O}_\alpha \rangle  \right] =0~.
\end{multline}

\begin{ack}
We would like to thank M.~Bianchi for stimulating
discussions and K.~Skenderis for clarifications regarding the standard
method and for helping us with the revised version. 
D.~M.\ is pleased to thank Universit\`a di Napoli ``Federico II''
and Universit\`a di Roma ``Tor Vergata'' for very kind hospitality
during completion of this work, as well as generous financial support.
D.~M.\ acknowledges partial support by PPARC
through SPG\#613. 
W.~M.\ acknowledges the financial support provided through the European 
Community's Human Potential Programme under contract HPRN-CT-2000-00131 
Quantum Spacetime, in which he is associated with INFN, LNF Frascati.
\end{ack}

\appendix
\section{Hamilton-Jacobi Method}
\label{HJmethod}
In this appendix, we shall summarize the Hamilton-Jacobi approach for a
system of Einstein gravity coupled to a scalar field and a vector field treated
in the St\"uckelberg formalism. Although the summary is sufficiently
self-contained for the purpose of this article, the reader is referred
to standard texts, \emph{e.g.}, \cite{MTW}, for a detailed description
of the method. 

The Hamiltonian treatment of gravity involves the time-slicing
formalism, which assumes that the bulk space-time manifold can be
globally foliated into hypersurfaces specified by a ``time''
coordinate.\footnote{For other cases, such as Taub-NUT spacetimes (see
\cite{Hawking:1998jf,Hawking:1998ct} and references therein), the
method must be extended.} 
  Of course, with Euclidean signature, there is no
distinction between time- and space-like directions, but the method
can be applied equally. As for notation, we adorn geometric bulk
quantities with a tilde and leave those belonging to hypersurfaces
unadorned. Greek indices, $\mu, \nu$, run from $0$ to $d$, latin
indices, $i, j$, from $1$ to $d$, and the index $r$ is often used
instead of the index $0$. 
Our conventions for the curvature tensor are $R^i{}_{jkl} = \partial_k
\Gamma^i{}_{jl} + \Gamma^i{}_{km} \Gamma^m{}_{jl} - (k\leftrightarrow
l)$, $R_{ij} = R^k{}_{ikj}$. 

For the applications of the present paper (see \cite{anatomy,Bianchi01b}),
we shall consider the following action:
\begin{equation}
\label{app:action}
\begin{split}
  S &= \int d^{d+1}x \sqrt{\tg} \left[ -\frac14\tR + \frac12
  \tg^{\mu\nu} \partial_\mu \phi \partial_\nu \phi +V(\phi)  \right] 
  + \frac12 \int d^d x \sqrt{g}\, H \\
  &\quad + \int d^{d+1}x \sqrt{\tg} \left[ \frac14 K(\phi) F_{\mu\nu}
  F^{\mu\nu} +\frac12 \M^2(\phi) B_\mu B^\mu \right]~.
\end{split}
\end{equation}
Here, $V(\phi)$ denotes a scalar potential, which has a local minimum
(stable fixed point) at $\phi=0$. 
The second integral in eqn.\ \eqref{app:action} is the Gibbons--Hawking
term, where $H$ is the trace of the second fundamental form of the
boundary hypersurface. This term is 
included in order to remove second derivatives with respect to the boundary
normal from the bulk integral. The vector sector contains a vector
field $A_\mu$ and the St\"uckelberg field $\alpha$, with $B_\mu =A_\mu
+\partial_\mu \alpha$ and $F_{\mu\nu}=\partial_\mu B_\nu - \partial_\nu
B_\mu$. 

Writing the bulk metric as 
\begin{equation}
\label{metric}
  \tg_{\mu\nu} = \begin{pmatrix} 
	n_i n^i +n^2  & n_j \\
	n_i           & g_{ij} 
	       \end{pmatrix}~,
\end{equation}
where $n$ and $n^i$ are called the lapse and shift functions,
respectively, the second fundamental form of $r=\text{const}$
hypersurfaces is given by 
\begin{equation}
\label{Hij}
  H_{ij} = -\frac1{2n} \left( \partial_r g_{ij} -\nabla_i n_j -\nabla_j
  n_i \right)~. 
\end{equation}

Using geometric identities, eqn.\ \eqref{app:action}
can be identically re-written as\footnote{Notice that $R$ denotes the
intrinsic curvature of $r=$const hypersurfaces.}  
\begin{equation}
\label{app:action2}
\begin{split}
  S &= \int d^{d+1}x \sqrt{g} n \left[ \frac14 (-R +H^i_j H^j_i -H^2) 
  + \frac12 \tg^{\mu\nu} \partial_\mu \phi \partial_\nu \phi +V(\phi)
  \right]  \\
  &\quad + \int d^{d+1}x \sqrt{g} n \left[ \frac14 K(\phi) F_{\mu\nu}
  F^{\mu\nu} +\frac12 \M^2(\phi) B_\mu B^\mu \right]~.
\end{split}
\end{equation}

The action \eqref{app:action2} is invariant under bulk diffeomorphisms
and gauge transformations, $\delta A_\mu = - \partial_\mu \delta
\alpha$. We shall gauge fix the quantities 
\begin{equation}
\label{gauge}
  n=1~, \quad n^i=0~, \quad A_r=0~,
\end{equation} 
so that their corresponding equations of motion will enter as
constraints into the Hamilton formalism,
\begin{align}
\label{eqmot_n}
  -\frac{\delta S}{\delta n} &= \Ham =0~, \\
\label{eqmot_ni}
  \frac{\delta S}{\delta n^i} &= \Ham_i =0~, \\
\label{eqmot_Ar} 
  \frac{\delta S}{\delta A_r} &= \G = 0~,
\end{align}
where (in gauge fixed form)
\begin{align}
\label{H}
  \Ham &= \frac14 \left( R + H^i_j H^j_i - H^2 \right) 
   +\frac12 (\partial_r \phi)^2 +\frac12 \M^2 B_r^2 +\frac12 K g^{ij}
   F_{ri} F_{rj} \\
\notag
  &\quad - \frac12 g^{ij} \partial_i \phi \partial_j \phi
   - \frac12 \M^2 B_i B^i - \frac14 K F_{ij} F^{ij} - V~, \\
\label{Hi}
  \Ham_i &= \frac12 \nabla_j ( \delta^j_i H - H^j_i) - \partial_r \phi
  \partial_i \phi -\M^2 B_r B_i - K g^{jk} F_{ij} F_{rk} \\
\intertext{and}
\label{G}
  \G &= \M^2 B_r + \nabla^i( K F_{ri} )~. 
\end{align}

The gauge fixed action reads 
\begin{equation}
\label{action_fixed}
\begin{split}
  S &= \int d^{d+1} x \sqrt{g} \left[ -\frac14 \left( R -H^i_j H^j_i +
  H^2\right) + \frac12 (\partial_r\phi)^2 +\frac12 g^{ij} \partial_i
  \phi \partial_j \phi + V \right. \\
  &\quad \left. + \frac12 \M^2 B_r^2 + \frac12 K g^{ij} F_{ri} F_{rj}
  +\frac12 \M^2 B_i B^i +\frac14 K F_{ij} F^{ij} \right]~,
\end{split}
\end{equation}
where $H_{ij} = -\frac12 \partial_r g_{ij}$, $B_r=\partial_r \alpha$
and $F_{ri} =\partial_r A_i$.

This gauge fixed system can be treated in Hamilton
language. The conjugate momenta are 
\begin{align}
\label{momenta_q}
  q^{ij} &= \frac1{\sqrt{g}} \frac{\delta S}{\delta(\partial_r
  g_{ij})} = \frac14 \left(g^{ij} H -H^{ij} \right)~, \\ 
\label{momenta_pi}
  \pi &= \frac1{\sqrt{g}} \frac{\delta S}{\delta(\partial_r \phi)} 
  = \partial_r \phi~, \\
\label{momenta_a}
  \pi_\alpha &= \frac1{\sqrt{g}} \frac{\delta S}{\delta(\partial_r
  \alpha)} = \M^2 \partial_r \alpha~,\\
\label{momenta_E}
  E^i &= \frac1{\sqrt{g}} \frac{\delta S}{\delta(\partial_r
  A_i)} = K g^{ij} F_{rj}~.
\end{align}
Expressed in terms of the momenta, the constraints \eqref{eqmot_n},
\eqref{eqmot_ni} and \eqref{eqmot_Ar} become 
\begin{align}
\label{con_H}
  \Ham &= 4 q^i_j q^j_i - \frac{4}{d-1} q^2 +\frac12 \pi^2 +
  \frac1{2\M^2} \pi_\alpha^2 +\frac1{2K} E_i E^i \\
\notag 
  &\quad +\frac14 R -\frac12 g^{ij} \partial_i \phi \partial_j \phi 
  - V  -\frac12 \M^2 B_i B^i -\frac14 K F_{ij} F^{ij} = 0~,\\
\label{con_Hi}
  \Ham_i &= 2 \nabla_j q^j_i - \pi \partial_i \phi -B_i \pi_\alpha -
  F_{ij} E^j =0~, \\
\label{con_G}
  \G &= \pi_\alpha + \nabla_i E^i =0~.
\end{align}
It is easy to realize that $\Ham$ coincides with the canonical
hamiltonian density. 

The bulk theory is defined on a bulk spacetime with a boundary at
$r=\rho$, where $\rho$ is a cut-off parameter. 
In the Hamilton--Jacobi formalism the momenta of the theory are obtained 
from the on-shell action $S$ as a functional of prescribed
boundary data, $g_{ij}(x,\rho)$, $\phi(x,\rho)$, $\alpha(x,\rho)$ and
$A_i(x,\rho)$, from 
\begin{align}
\label{q_S}
  q^{ij} &= \frac1{\sqrt{g}} \frac{\delta S}{\delta g_{ij}}~,\\
\label{pi_S}
  \pi  &= \frac1{\sqrt{g}} \frac{\delta S}{\delta \phi}~, \\
\label{pia_S}
  \pi_\alpha &= \frac1{\sqrt{g}} \frac{\delta S}{\delta \alpha}~, \\
\label{E_S}
  E^i &=  \frac1{\sqrt{g}} \frac{\delta S}{\delta A_i}~,
\end{align}
where the variation is with respect to the boundary data. Therefore equations
(\ref{con_H} - \ref{con_G}) become constraints to be satisfied by $S$.


\end{document}